\documentclass{kapproc} 
\usepackage{epsfig}

\kluwerbib

\begin{document}

\articletitle{High energy emission\\ from microquasars}

\author{Rob Fender and Tom Maccarone}
\affil{Astronomical Institute `Anton Pannekoek', University of Amsterdam\\
Kruislaan 403, 1098 SJ Amsterdam, The Netherlands}
\email{rpf@science.uva.nl and tjm@science.uva.nl}

\begin{abstract}
The microquasar phenomenon is associated with the production of jets
by X-ray binaries and, as such, may be associated with the majority of
such systems. In this chapter we briefly outline the associations,
definite, probable, possible, and speculative, between such jets and
X-ray, $\gamma$-ray and particle emission.
\end{abstract}


\section*{Introduction: what is a microquasar ?}

The answer to the above question depends somewhat on your viewpoint,
but ours is the following: a `microquasar' is an X-ray binary (XRB)
which produces jets. Figure 1 is a sketch of an XRB, presenting the major
physical components and sites of emission in such systems. Currently
about 250 XRBs are known in our galaxy (comprehensively catalogued in
Liu, van Paradijs \& van den Heuvel 2000, 2001), possibly representing
an underlying population of $\geq 1000$ objects, depending on the
value of any low-luminosity cut off to the distribution (e.g. Grimm,
Gilfanov \& Sunyaev 2002). 

\begin{figure}[h]
\centerline{\epsfig{file=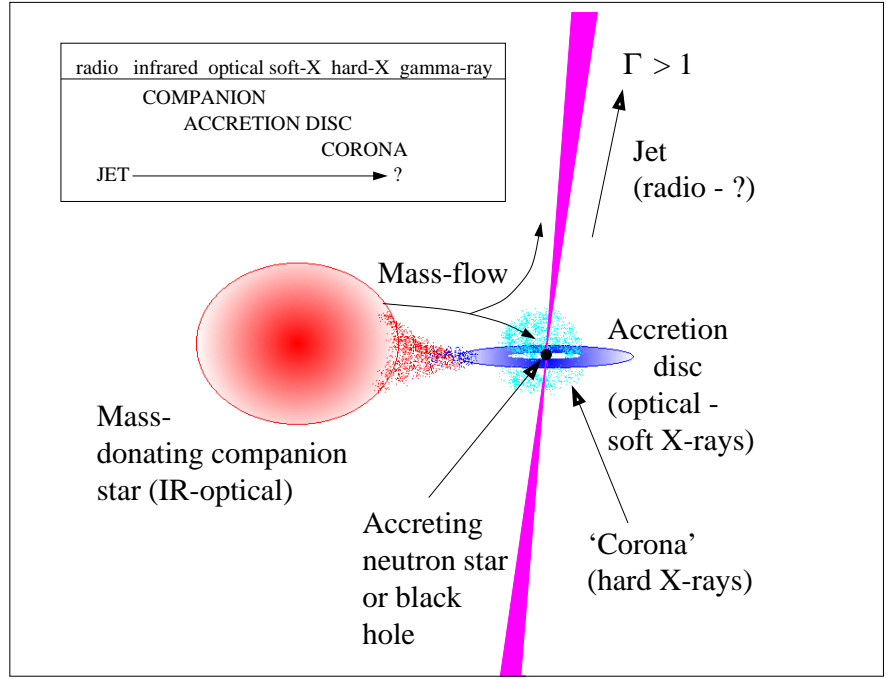, width=9cm, angle=-90}}
\caption{
Sketch illustrating our current concept of the physical components and
sites of emission in an X-ray binary system. A compact object (neutron
star or black hole) accretes material from a binary companion, and the
potential accretion energy is released in the form of a combination of
high-energy emission (UV / X-rays / $\gamma$-rays) and mechanical
energy in an outflow (which itself may be the site of some high-energy
emission).
}
\end{figure}

Figure 2 presents the spatial distribution of known XRBs: these can be
separated into two populations: the low-mass XRBs (where `low mass'
refers to the binary companion star) which are believed to be older,
and are concentrated near the Galactic bulge, and the younger
high-mass XRBs concentrated in the spiral arms. In general, the low
mass X-ray binaries transfer mass to their companions through Roche
lobe overflow, while the high mass X-ray binaries transfer mass
through strong stellar winds.  For the purposes of jet formation the
mass transfer mechanism does not seem to be important, and we can
consider Fig 2 as demonstrating the roughly uniform distribution of
XRBs with mass in the galaxy.

Based on our definition of a microquasar, about 15\% of the Milky
Way's X-ray binaries (including nearly all of its X-ray binaries
thought to contain a black hole) definitely fall into this
class. However, in our opinion, jet production is in fact likely to be
common for up to 70\% of X-ray binaries (see arguments in Fender
2004). In this case it is perhaps more appropriate to consider
`microquasar' as a {\em phenomenon} associated with XRBs, rather than
considering microquasars to be some unusual subset of objects.

The jets from XRBs appear to come in two types. Firstly, in `hard'
X-ray states -- typically observed for bolometric X-ray luminosities
$\leq 5$\% Eddington (Maccarone 2003) -- a steady, self-absorbed
outflow appears to be ubiquitous in both black hole (BH; e.g. Fender
2001) and neutron star (NS; Migliari et al. 2003a) XRBs. These jets
appear to be self-absorbed (based upon their flat or inverted radio
spectra: corresponding to spectral index $\alpha \geq 0$ where the
relation between flux density and frequency is given by S$_{\nu}
\propto \nu^{\alpha}$). They are also strongly coupled to the X-ray
emission from these systems, evident in an apparently universal
correlation between radio and X-ray luminosities of the form

\[
L_{\rm radio} \propto L_{\rm X}^b
\]

where $b \sim 0.7$ (Corbel et al. 2003; Gallo, Fender \& Pooley
2003). Based on the limited scatter around a single function of this
form, Gallo, Fender \& Pooley (2003) concluded that the bulk Lorentz
factor ($\Gamma = [1-\beta^2]^{0.5}$, where the velocity $v=\beta c$) is
probably less than 2.

\begin{figure}
\centerline{\epsfig{file=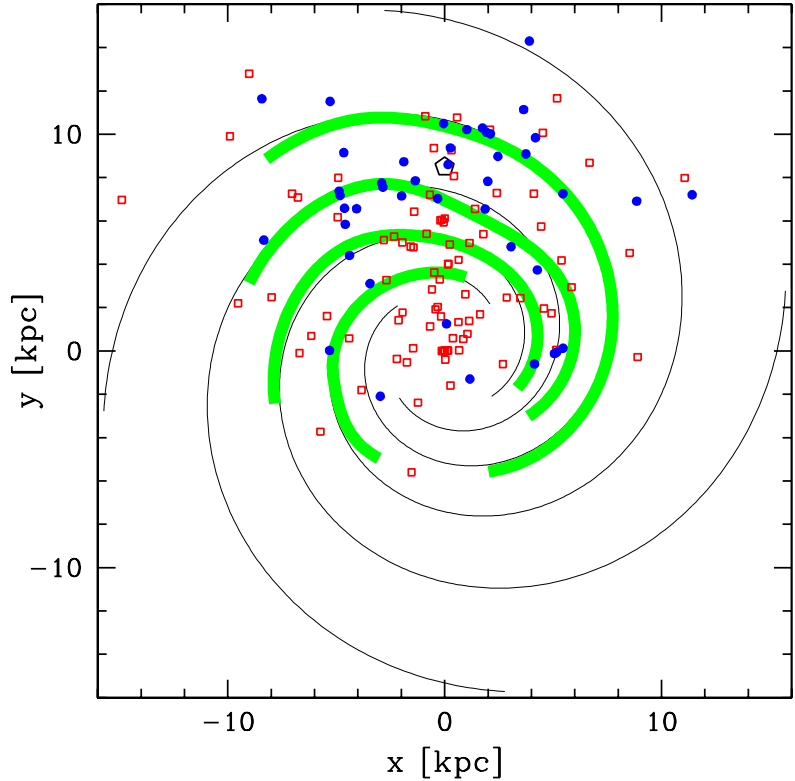, width=11cm, angle=0}}
\caption{
The relatively uniform distribution of known XRBs within our
galaxy. The Sun is located at (0,8.5). From Grimm et al. (2002).  
}
\end{figure}

\begin{figure}[h]
\centerline{\epsfig{file=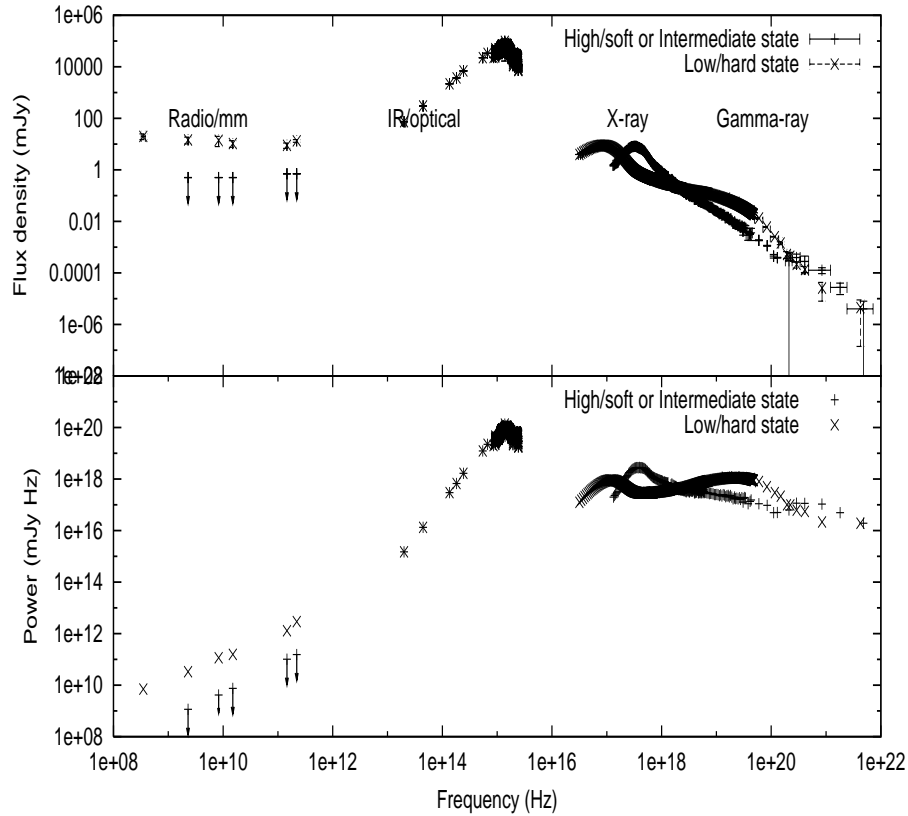, width=12cm, height=12cm}}
\caption{Broadband spectra -- over 14 decades in frequency -- of
  Cygnus X-1 in the `low/hard' and softer (`high/soft' or
  `intermediate') states, in flux density and spectral power
  representations. The radio--mm emission is significantly stronger in
  the `low/hard' state, which has an X-ray spectrum which peaks in
  power at $\sim 100$ keV. The infrared--optical regime is dominated
  by the OB companion star, whose emission (of course) does not change
  between states. Note that despite the significantly altered radio
  and X-ray spectra, $\gamma$-ray detections at $\geq 1$ MeV are
  comparable in both X-ray states. From Tigelaar et al. (2003).}
\label{nu_bw}
\end{figure}

At slightly higher X-ray luminosities, the X-ray spectra of XRBs
softens. In the `high/soft' states of BH XRBs the radio emission is
dramatically `quenched', probably indicating suppression of jet
formation (Tananbaum et al. 1972; Fender et al. 1999; Gallo, Fender \&
Pooley 2003). There is a hint that similar behaviour may be exhibited
by NS XRBs (Migliari et al. 2003). At higher luminosities (but see
Homan et al. 2001) the `intermediate' or `very high' state of BHs --
{\em or the rapid transition to this state} -- is associated with the
formation of discrete, powerful jets. These jets are likely to have
bulk Lorentz factors $\Gamma > 2$ and so are different to the steady
jets in the low/hard state (for further discussion see Fender
2004). At the highest luminosities NS XRBs also display powerful,
highly relativistic jets (e.g. Fomalont et al. 2001; Fender et
al. 2003).  Overall the BH and NS XRBs appear to show analogous
patterns of jet formation as a function of (apparent) bolometric
luminosity, although the NS XRBs are a factor 10--100 less `radio
loud' than the BH XRBs (Fender \& Kuulkers 2001; Migliari et
al. 2003).

The high/soft state is generally thought to be dominated by a
geometrically thin, optically thick accretion disk (e.g. Shakura \&
Sunyaev 1973).  As such, it may be incapable of extracting rotational
energy efficiently, either from itself or from the central black hole
(e.g. Livio, Ogilvie \& Pringle 1999; Meier 2001; Livio, King \&
Pringle 2003), as thin disks are expected to have weaker poloidal
magnetic fields than thick disks.  That Galactic black holes, neutron
stars, and more recently, active galactic nuclei (Maccarone, Gallo \&
Fender 2003) all show suppression of jet formation in the $\sim$2-10\%
of Eddington luminosity range where high/soft states are seen bolsters
the case for this theoretical picture.

Figure 3 demonstrates the broad band spectra of hard and soft X-ray
states in the BH XRB Cygnus X-1, clearly indicating the `quenching' of
the radio and mm emission in the `soft' X-ray state. In the context of
this chapter it is interesting to note that the source is detected at
MeV energies in both states (see McConnell et al. 2002 for a more
detailed discussion). 

\section*{X-ray emission from jets ?}

In the following section we shall discuss the observational evidence
and theoretical interpretations / speculations regarding the emission
of X-rays from the two types of XRB jets outlined above.

\subsection*{X-rays from steady jets ?}

\begin{figure}
\centerline{\epsfig{file=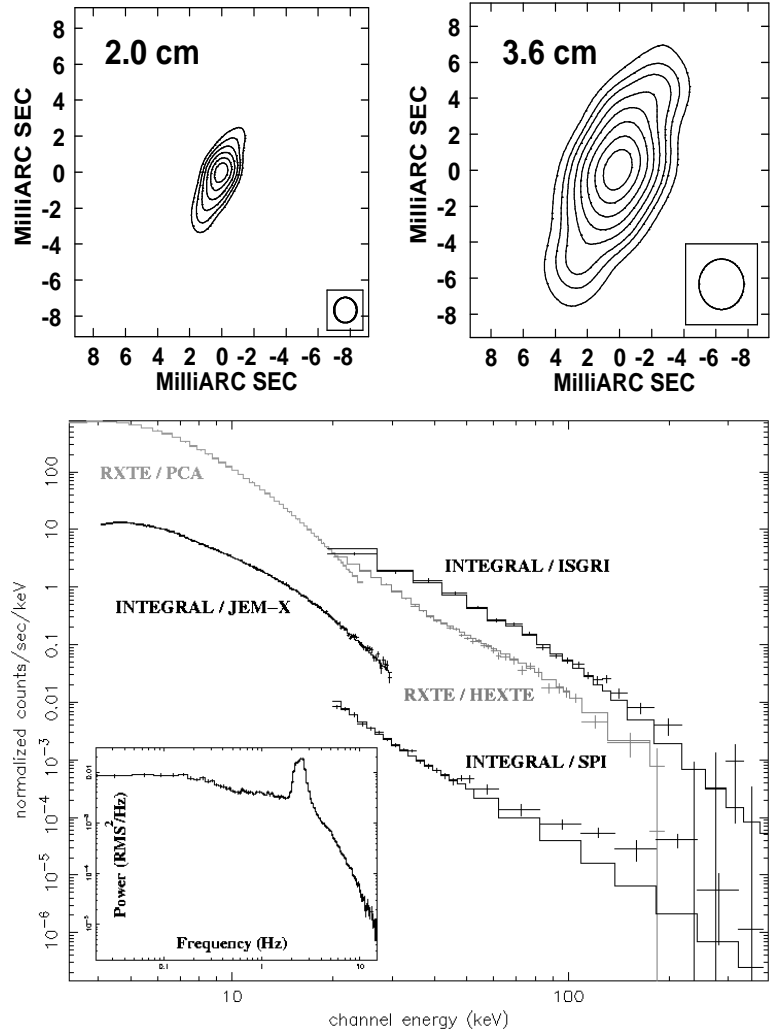, width=16cm, angle=270}}
\caption{High-energy emission from GRS 1915+105 in a
  steady-jet-producing `plateau state'. The top panel shows VLBA
  observations of GRS 1915+105 revealing a steady, compact, jet during
  a period of steady, hard X-ray emission, commonly referred to as a
  plateau. The lower panel shows the X-ray spectrum of the source
  during this period, as observed with RXTE and INTEGRAL.  The inset
  shows the X-ray power spectrum as measured by RXTE. From Fuchs et
  al. (2003).  }
\label{fuchs}
\end{figure}

It should be stated from the start that there is no direct evidence
that the steady jets associated with hard X-ray states are the sites
of any of the observed X-ray emission. What is clear is that there is
a strong coupling between radio and X-ray luminosities (as noted
above), and that this therefore implies a strong coupling between the
jet and accretion flow. Figure {\ref{nu_bw}} demonstrates the dramatic
change in the radio--mm component of the spectrum of a BH XRB (Cygnus
X-1) correlated with changes in the X-ray spectrum. Note that in the
low/hard state of this source a jet has been directly imaged with
radio VLBI (Stirling et al. 2001). Figure {\ref{fuchs}} shows the
spectrum of the powerful XRB jet source GRS 1915+105 in a `plateau'
state (the nearest it gets to the `standard' low/hard state), together
with radio VLBI images made contemporaneously which reveal a steady
compact jet. It is clear that when such steady jets are present there
are a lot of hard X-rays. It is possible that this X-ray component
{\em does} originate in the base of the jet, via optically thin
synchrotron emission (e.g. Markoff, Falcke \& Fender 2001) but this is
not the widely accepted interpretation, which is instead that the
X-rays arise via Comptonisation of softer photons in a hot ($\sim 100$
keV) plasma (e.g. Thorne \& Price 1975; Zdziarksi et al. 2003).  

\subsection*{X-rays from transient relativistic jets ?}

\begin{figure}
\centerline{\epsfig{file=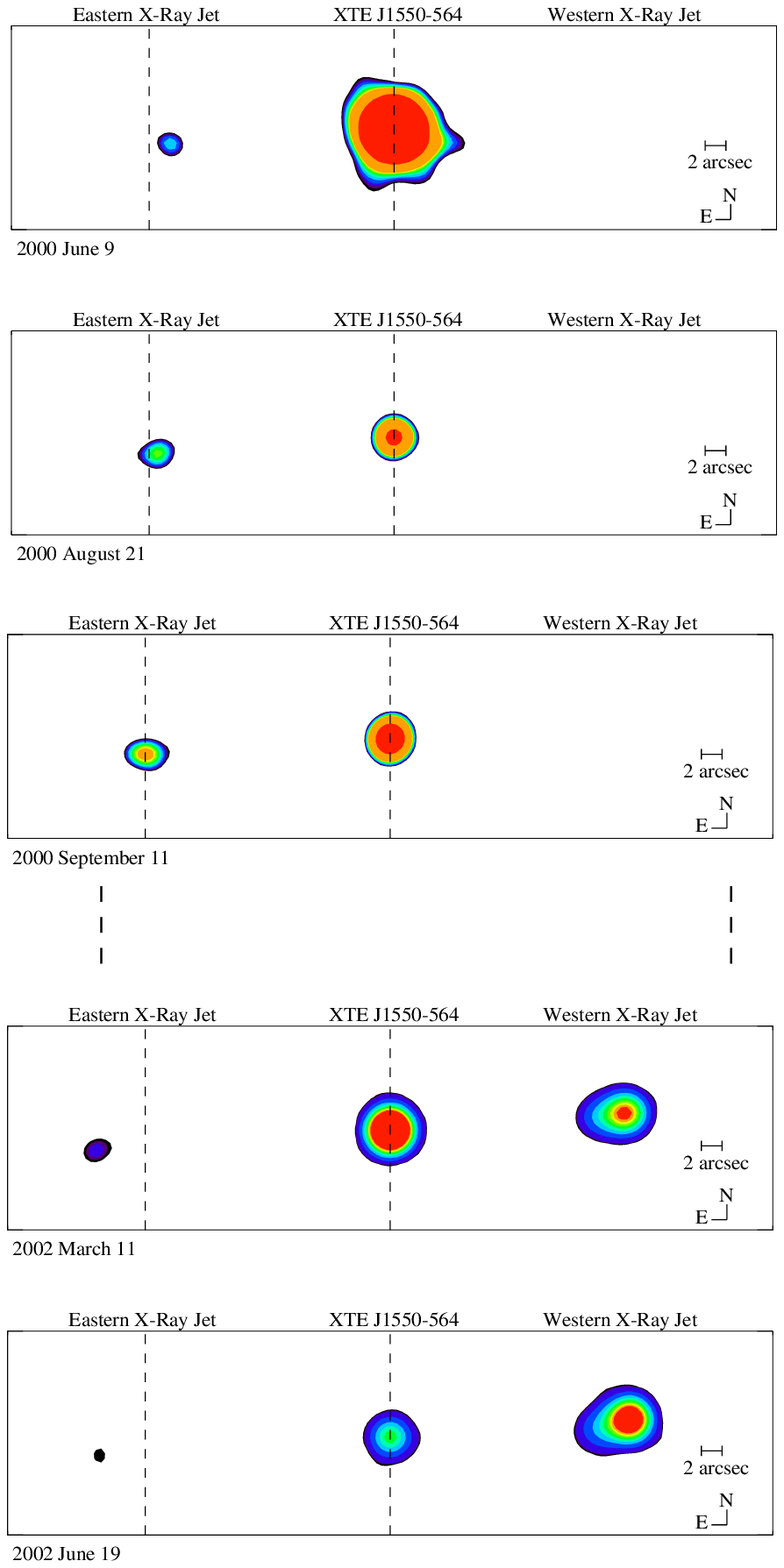, width=10cm, angle=0}}
\caption{X-ray emission from large-scale jets produced during an
  outburst of the black hole transient XTE J1550564. Adapted from
  Corbel et al. (2002).  Under minimum energy conditions, the
  leptons radiating in the soft X-ray band should have TeV energies.}
\label{1550}
\end{figure}

This question is much easier to answer, with an unambigous {\em
yes}. Figure {\ref{1550}} presents {\em Chandra} X-ray images of moving
large-scale jets from the transient BH XRB XTE J1550-564, two years
after a major outburst (Corbel et al. 2002; Kaaret et al. 2003;
Tomsick et al. 2003; see also Wang, Dai \& Lu 2003). While XTE
J1550-564 was very bright, there was nothing outstandingly unusual
about it, and so we may conclude that large-scale X-ray emitting jets
are probably rather commonly associated with XRB outbursts. The
radio-through-X-ray spectrum of these moving jets can be fit by a
single power law with spectral index $\alpha = -0.660 \pm 0.005$,
entirely consistent with optically thin synchrotron emission.
Applying minimum energy arguments (see e.g. Longair 1994), we can
derive a magnetic field in the jets of 0.3mG. This in turn implies
that the leptons emitting in the soft X-ray band have been accelerated
to TeV energies, presumably by an interchange of bulk kinetic energy
to particles via shocks. Large-scale X-ray jets are also associated
with the unusual X-ray binary SS 433 (Brinkmann, Aschenbach \& Kawai
1996; Migliari, Fender \& Mendez 2002) and the transient 4U 1755-33
(Angelini \& White 2003).  We also note that extended jets represent
an additional possible mechanism for producing the elongated X-ray
sources seen in deep Chandra images of the Galactic Center region,
previously suggested to be predominantly pulsar wind nebulae (Lu, Wang
\& Lang 2003).

\section*{High-energy $\gamma$-ray emission}

For the purposes of this chapter, we refer to any $\gamma$-rays above 30
MeV as high energy (HE) $\gamma$-rays, and $\gamma$-rays above 100 GeV
as VHE $\gamma$-rays.  In this way, HE $\gamma$-rays are restricted to
be photons which nearly all models have associated with some kind of
high velocity relativistic shocks, generally in a jet.  The VHE
$\gamma$-rays are defined to be only those photons which can be
detected only through ground-based $\gamma$-ray observatories.  We
will concentrate on the emission detected from jets here, rather than
the emission from systems which are known not to have jets, such as
the X-ray pulsars (for example, Cen X-3 - see Chadwick et al. 2000).

\subsection*{Observations}

There have been numerous claims of detection of HE $\gamma$-rays from
X-ray binaries, but to date, with three that are fairly convincing.
These detections have come primarily from EGRET (see Hartman et
al. 1999 for the EGRET catalog, and the text below for a discussion of
the individual X-ray binary jet sources), although a few detections
have been claimed from the ground-based Cerenkov telescopes.  There
have also been suggestions that many of the EGRET unresolved sources
in the Galactic plane are X-ray binaries (e.g. Kaufman Bernad\'o,
Romero \& Mirabel 2002) or other accreting stellar mass compact
objects with jets (e.g. Armitage \& Natarajan 1999; Punsly et
al. 2000).

\subsubsection*{Cygnus X-3}
The first X-ray binary with jets suggested to emit TeV $\gamma-$rays
is Cygnus X-3 (Samorski \& Stamm 1983; Chadwick et al. 1985).  The
detection is bolstered by the fact that the TeV $\gamma-$rays are
strongest during the X-ray maxima in the 4.8 hour orbital period
(Chadwick et al. 1985).  Other groups have failed to detect Cygnus X-3
during the X-ray maximum (e.g. O'Flaherty et al. 1992), but this may
be a result of variability.

Additionally, a 12.6 msec periodicity has been reported by multiple
groups, although the period measurements are quite precise and seem
not to be exactly consistent with one another (Chadwick et al. 1985;
Brazier et al. 1990; Gregory et al. 1990).  On the other hand, these
pulse periods are based on rather small numbers of photons.
Furthermore, the statistical singificance of the pulse detection has
been debated on the ground that the calculations of the signficance
level have ignored the DC component of the emission (Protheroe 1994).

The association of the periodicity with an actual pulsar (i.e. a
rotating neutron star) is a bit more troubling.  Firstly, jets have
been observed from Cygnus X-3 (in fact, the brightest radio jets of
any X-ray binary in the Galaxy), but have not been observed from any
other accreting X-ray pulsar; it is believed that the strong magnetic
field in the X-ray pulsars may suppress jet formation (e.g. Fender \&
Hendry 2000).  Furthermore, the fact that Cyg X-3 at its peak is has
the highest ratio of radio to X-ray flux of any Galactic X-ray binary
indicates that it probably contains a black hole accretor, since the
neutron stars typically fall 10-100 times below the black holes in
their ratios of radio to X-ray flux.  Additionally, the 12.6 msec
period has never been detected in other wavelengths, although this may
be partly because of scattering through an extremely dense wind
environment.  It is believed on dynamical grounds that Cygnus X-3
might contain a black hole (see e.g. Schmutz et al. 1996; Hanson,
Still \& Fender 2000).

\subsubsection*{LS 5039}
Another strong case of an X-ray binary emitting HE $\gamma$-rays can
be made for LS 5039.  It shows persistent radio jets with a size scale
of $\sim 10^{14}$ centimeters, and is positionally coincident with an
EGRET source.  The $\gamma$-ray emission also seems to be persistent
(Paredes et al. 2000 and references within).  The very high ratio of
$\gamma$-ray to radio power, in combination with the rather large size
scale of the radio emitting region led Paredes et al. (2000) to
suggest that that the dominant emission process is inverse
Comptonization with seed photons dominated by the bright mass-donating
star.

\subsubsection*{LS I +61 303}
Relatively strong evidence exists for $\gamma$-ray emission from this
source as well.  There is a positional coincidence with the COS B and
EGRET source 2 CG 135+01 (Gregory \& Taylor 1978; Kniffen et
al. 1997).  This source is highly variable in the radio and X-ray
bands, as it is a B[e] accreting binary with a highly eccentric orbit.
However, the $\gamma-$rays do not show any clear sign of variability,
let alone variability correlated with that in the radio (Kniffen et
al. 1997).  Still, there are no other strong candidate sources within
the EGRET error boxes.

\subsubsection*{Unidentified EGRET sources in the galactic plane} 

Almost half of the unidentified EGRET sources are within 10 degrees of
the Galactic plane (Hartmann et al. 1999), so it seems likely that a
large fraction of the sources are Galactic in origin.  While recent
follow-ups suggest that some of the previously unidentified sources
may be young pulsars (Halpern et al. 2001; D'Amico et al. 2001; Torres
et al. 2001), the spatial distribution seems more consistent with that
of molecular clouds (e.g. Bhattacharya et al. 2003), and the follow-up
of LS 5039 as the counterpart of a previously unidentified EGRET
source suggests that at least some fraction of these sources may be
X-ray binaries.

\section*{Predictions}
Numerous models make different predictions for the VHE $\gamma$-ray
production.  Below we discuss these models briefly.

\section*{Synchrotron self-Comptonization models}

The simplest mechanism invoked for explaining the VHE gamma-rays seen
in blazars are those which invoke synchrotron self-Compton radiation.
In such a model, the same electrons produce two broad peaks in the
spectrum.  A low energy peak comes from synchrotron radiation, and a
higher energy peak comes from Compton upscattering (e.g. Jones, O'Dell
\& Stein 1974).  The photon energies of the two peaks depend on the
electron energy distribution and on the magnetic field.  The ratio of
the luminosity of synchrotron component to the luminosity of the
Compton component is equal to the ratio of the magnetic energy density
to the photon energy density.

\begin{figure}[h]
\centerline{\epsfig{file=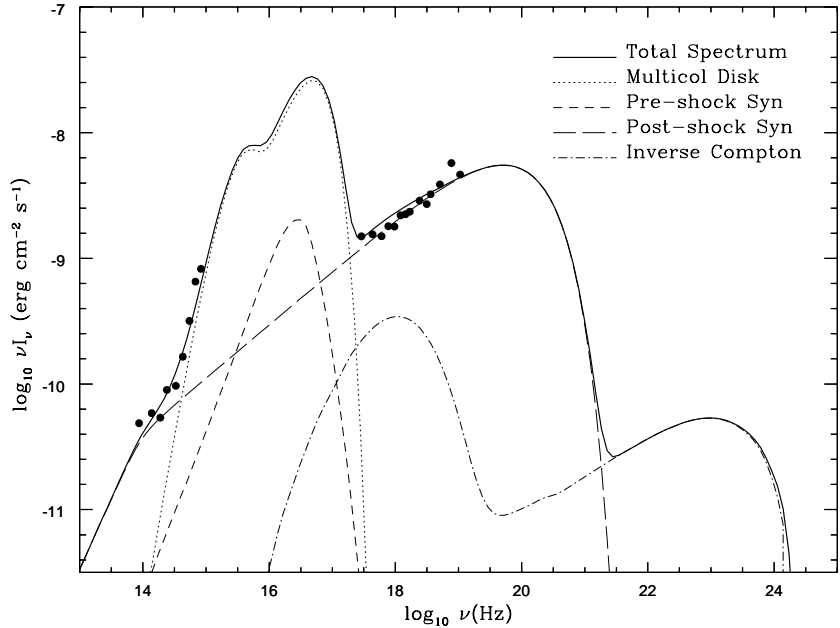, angle=270, width=12cm}}
\caption{Figure 3 (adapted from Markoff et al. 2003), which shows the jet model
fits to the 1981 radio-through-X-ray data for the bright low/hard state of
GX 339-4 and predictions for the inverse Compton emission from this model.}
\label{Markoff03}
\end{figure}

A recent set of models has suggested that in some X-ray binaries, the
hard X-rays may be dominated by synchrotron emission from a
relativistic jet.  Two strong examples are XTE J1118+480 (Markoff,
Falcke \& Fender 2001) and GX 339-4 (Markoff et al. 2003), although in
this general framework some other sources show X-ray emission
dominated by the Compton component of the jet.  Since the TeV blazars
all have X-ray emission dominated by synchrotron emission from the
jet, these seem to be ideal candidates for observing TeV emission from
an X-ray binary.  In fact, for the case of GX 339-4, the synchrotron
self-Compton spectrum has been presented for the fit to the brightest
low/hard state for which there is good multi-wavelength data; the
predicted $\gamma$-ray fluxes are below the sensitivity of $\sim$ few MeV
instruments such as those aboard INTEGRAL, but above the detection
limits anticipated from GLAST, and near the one night detection limits
for STACEE and HESS (see Figure \ref{Markoff03}, reproduced from Markoff et
al. 2003).  The magnetic field predicted for X-ray binaries may be
higher than that inferred for blazars, which means that despite the
fact that the hard X-rays from GX 339-4 extend to energies seen only
in Mkn 501 in the blazars, the electron energies and hence the highest
energies of the Compton scattered photons are a factor of about 100
lower than seen in Mkn 501.

\section*{External Comptonization models}

The next step up in complication from the synchrotron self
Comptonization models is external Comptonization.  In such a picture,
there is an additional source of seed photons to the Comptonizing
region aside from the synchrotron photons.  In the context of AGN,
this additional source of photons is often assumed to be the accretion
disk (Dermer \& Schlickeiser 1993) or the broad line region
(e.g. Sikora, Begelman \& Rees 1994).  The broad line region is
usually located not too far from the active region of the jet, and it
is further from the central engine than is the active region of the
jet, so the jet's motions will be toward the BLR.

In the context of X-ray binaries, a possible important source of
external photons is the mass donating star.  External Comptonization
in the relativistic jet has been suggested to explain the hard tail
seen by COMPTEL (McConnell et al. 2002) for Cygnus X-1
(Georganopoulos, Aharonian \& Kirk 2002; Romero, Kaufman Bernado \&
Mirabel 2002).  It has also been suggested as an explanation for the
possible association of LS 5039 with an EGRET source.  In both cases,
the mass donor is a high mass star; for the low mass X-ray binaries,
the photon fields of the donors are likely to be too weak to provide
significant seed photon populations.  It may also be that external
Comptonization will be more important if the jet is not perpendicular
to the binary's orbital plane; this seems to be the case whenever good
measurements of both the jet and orbital plane inclination angles are
measurable, and seems to be especially likely to occur for high mass
X-ray binaries which have short system lifetimes and hence cannot
change the spin axis of the mass accretor (Maccarone 2002).  A
particularly promising candidate for this misalignment effect is V4641
Sgr, which has an extremely strong misalignment, and also is a
microblazar (see e.g. Orosz et al. 2001).  Attempts to fit jet models
incorporating the external photons fields from the accretion disc and
from the companion star are underway (Markoff \& Maccarone, in prep).

\begin{figure}
\centerline{\epsfig{file=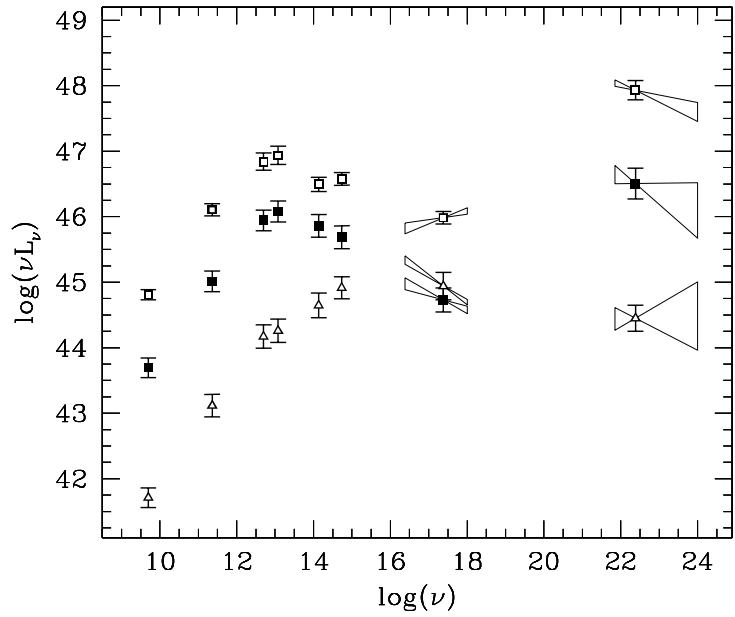, width=10cm}}
\caption{Figure 10 from Fossati et al. 1998, which shows how blazars
shift to lower energy peaks with higher fractions of their total power
in the Compton component at higher luminosities.  The open boxes
represent flat spectrum radio quasars (the brightest sources), the
filled boxes represent the 1 Jy BL Lac sample (intermediate brightness
sources), and the traingles represent a fainter sample of BL Lac
objects.  Note also, the similarity with Figure \ref{Markoff03}, apart
from the accretion disk's contribution in GX 339-4, which is
relatively stronger than in the blazars because the jet is not beamed
towards the observer in GX 339-4.}
\label{fossati}
\end{figure}

In general, there are two key effects of external Comptonization.  The
first is that the ratio of luminosity in the Compton component to that
in the synchrotron component is boosted, since
$L_{synch}/L_{Compt}=U_{B}/U{ph}$ (see e.g. Rybicki \& Lightman 1979).
The second is that the cooling rate is boosted by the addition of the
additional soft photons, pushing the Compton and synchrotron peaks to
lower energies.  Evidence of this phenomenom in AGN is probably seen.
The brightest blazars presumably have the accretion disks and broad
line regions which contribute the most flux.  They also have the most
$\gamma-$ray dominant spectra and the lowest peak energies for both
the synchrotron and Compton components of the spectrum (e.g. Fossati
et al. 1998 - see Figure \ref{fossati}).  It is also noted that a pure
synchrotron self-Comptonization scenario could also explain the
observed correlations if there is a systematic variation in magnetic
field strengths, with the brightest blazars have the weakest magnetic
fields (Fossati et al. 1998).  This may have important implications
for detecting $\gamma-$ray emission from X-ray binaries, as well.

\section*{Hadronic jet models}
The three putative HE/VHE $\gamma$-ray sources among the X-ray
binaries are all high mass X-ray binaries with strong stellar winds in
addition to the strong photon fields.  Given that there will be dense
matter fields in the region where the jet exists, there should also be
strong collisions between the jet and the matter from the stellar
wind.  The collisions will lead to pion decays, and hence to the
production of $\gamma$-rays and neutrinos (Romero et al. 2003).  A
potential problem with this model is that the jets may be pair
dominated in the low Eddington fraction steady jet systems such as LS
5039 by analogy with AGN.  There is some evidence that the FR I jets
which are analogous to the low/hard state steady jets (Meier 1999;
Maccarone, Gallo \& Fender 2003) may be pair dominated (e.g. Reynolds
et al. 1996); if confirmed, then this model would be ruled out.  The
strong neutrino flux predictions may be testable even sooner.

\section*{Gamma-ray lines}

For the most part, $\gamma$-ray lines have been seen from diffuse
sources where nuclear reactions are occuring, rather than from point
sources.  However, there is at least one instance of a measurement of
a $\gamma$-ray line from an X-ray binary - that seen in Nova Mus 1991,
at 481$\pm$22 keV (Gilfanov et al. 1991). Two possible interpretations
of this emission line have been suggested - redshifted annihilation,
possibly from a recombination of pairs in a pair-dominated
relativistic jet (e.g. Kaiser \& Hannikainen 2002) or a $^7$Li
re-combination line at 478 keV, possibly related to a collision
between a mis-aligned relativistic jet and the mass donor (Butt,
Maccarone \& Prantzos 2003).  That the mass donor in Nova Mus 1991
shows a lithium excess (Martin et al. 1996) lends some support to the
idea that the spectral line is a Li line.  Further support may be lent
by the observational and theoretical evidence for the association of
jets ejected at high luminosities (like that in Nova Mus 1991) with
the FR II AGN jets (Meier 1999; Maccarone, Gallo \& Fender 2003).  The
energetics of FR II jets seem to suggest that they must be baryon
dominated (Celotti \& Fabian 1993).  The ubiquity of lithium excesses,
on the other hand is a point against the jet-star interaction scenario
for producing lithium during the outbursts.

\section*{Why are there so few sources, and how do we find more?}
Relatively few phenomena have been seen clearly in active galactic
nuclei and not in X-ray binaries.  Very high energy gamma-ray emission
seems to be one of these.  The question thus arises about whether this
is because of a physical difference between the two types of systems
or due to observational selection effects.  At the present time, all
the associations of high energy $\gamma$-rays with X-ray binaries
remain speculative.  On the other hand, there are many confirmed EGRET
blazars and a handful of TeV $\gamma$-ray AGN.  There are numerous
possible reasons for this:

\subsection*{Lower radio luminosities?}
One of the keys to identifying the EGRET blazars with their
counterparts at other wavelengths is the presence of a strong radio
source (typically above 1 Jy) within the error box.  It has been shown
that the ratio of radio to X-ray power for AGN is systematically much
higher than for X-ray binaries due to a mass term in the fraction of
the total jet power that comes out at a given radio frequency
(e.g. Heinz\& Sunyaev 2003; Merloni, Heinz \& Di Matteo 2003; Falcke,
K\"ording \& Markoff 2003).  Therefore, many of the EGRET unidentified
sources may be X-ray binaries in the Milky Way which have not been
associated with counterparts at other wavelengths as easily as have
the high latitude EGRET sources.

\subsection*{Rapid variability?}
It is generally believed that the variability timescales for accreting
objects vary linearly with the mass, as the mass is the most important
size scale for such systems.  Taking as a template, for example, the
durations of the TeV flares from Mkn 421 and Mkn 501, which are
typically a few months, and the estimated masses of the black holes in
these systems (typically greater than $10^8$M$_\odot$ - see
e.g. Barth, Ho \& Sargent 2003), we find that the peak $\gamma$-ray
emitting time period would be of order one second or less for a 10
$M_\odot$ black hole.  In reality, the situation may not be so bad;
the X-ray binaries are much closer than the blazars and hence are
brighter in all other wavelengths but radio.  They might thus be
observable far deeper in the outburst than the blazars.  The real
problem is likely to be one of duty cycle; the duration for which
X-ray binaries emit radiation is likely to be rather short because the
evolution of the outburst cycle is quite rapid.  Without excellent
sampling in the $\gamma$-rays over the peak of the outburst cycle, it
seems unlikely that the short period where the VHE $\gamma-$rays are
emitted will coincide with the observations.

\subsection*{Beaming requirements?}
The blazars from which we have seen VHE $\gamma-$rays are all highly
beamed.  Given that the probability of a seeing a source with Doppler
factor $\delta$ is roughly $\delta^{-2}$ and that there are only a few
hundred X-ray binaries in the Milky Way, the number of highly beamed
sources is quite small.  There are, however, a few sources that do
show strong evidence for being highly beamed - Cygnus X-3 which shows
a strong jet and no counterjet on small distance scales (see
e.g. Mioduszewski et al. 2001), and V4641 Sgr (Hjellming et al. 1999;
Orosz et al. 2001) and Cir X-1 (Fender et al. 2003), which show proper
motions corresponding to apparent velocities in excess of 15$c$.
Cygnus X-3 {\it is} one of the strongest candidates for having shown
VHE $\gamma$-ray emission among the Galactic source.  The outbursts of
V4641 Sgr have been extremely rapid, meaning that the response time of
current TeV observatories may have been insufficient.  We are not
aware of any attempt to observe Cir X-1 in the VHE $\gamma$-rays near
its radio peaks.

\subsection*{Magnetic field differences?}

Assuming that the total kinetic luminosity injected into the jet is
proportional to the black hole mass, and that this is injected on a
timescale proportional to the black hole mass, and that the magnetic
field is in equipartition with the particle energy, then the magnetic
field should be proportional to $M^{-0.5}$.  A higher magnetic field
means that (1) the synchrotron component will be stronger relative to
the Compton component of the jet spectrum and (2) as discussed in the
context of the Markoff et al. (2003) model for GX 339-4, even given
synchrotron X-rays, the particle energies will not need to be as high,
and there may not be TeV electrons present.

\subsection*{Poorer observational coverage?}
It is rather difficult to assess the role played by different
observational strategies in studying AGN and X-ray binaries.  Most of
the VHE $\gamma-$ray observatories have pointed instruments which
observe only a small field of view.  The notable exception is MILAGRO,
which has a much poorer sensitivity than the other instruments
(although it might be sufficient for detecting a highly beamed event
at high instrinsic luminosity from a nearby part of the Galaxy).  This
underscores the importance of publishing detailed upper limits
including the times when the observations were made.

\section*{Conclusions}

We have provided a brief overview of the X-ray and $\gamma$-ray
emission from X-ray binaries with jets -- also known as
`microquasars'. X-ray emission is unambiguously associated with
jet-ISM interactions following major transient outbursts. It is
certainly strongly coupled to the jet in steady states, although the
nature of the coupling and the site of emission of the X-rays is not
conclusively established. There are stong reasons, both directly from
observations and by analogy with active galactic nuclei, that high
energy $\gamma$-ray emission may also be associated with microquasars,
particularly from those sources with jets inclined close to our line
of sight (`microblazars'). New $\gamma$-ray observatories may well
find X-ray binaries to be an important contributor to the $\gamma$-ray
emission and particle acceleration within our Galaxy.

\end{document}